# Non-representative quantum mechanical weak values

B. E. Y. Svensson

*Theoretical High Energy Physics,Department of Astronomy and Theoretical Physics, Lund University, Sölvegatan 14, SE-22362 Lund, Sweden*

E-mail: Bengt_E_Y.Svensson@thep.lu.se

**Abstract**

The operational definition of a weak value for a quantum mechanical system involves the limit of the weak measurement strength tending to zero. I study how this limit compares to the situation for the undisturbed (no weak measurement) system. Under certain conditions, which I investigate, this limit is discontinuous in the sense that it does not merge smoothly to the Hilbert space description of the undisturbed system. Hence, in these discontinuous cases, the weak value does not represent the undisturbed system. As a result, conclusions drawn from such weak values regarding the properties of the studied system cannot be upheld. Examples are given.

## I. INTRODUCTION

The concept of a quantum mechanical *weak value* resulting from *weak measurement* followed by *postselection* has attained much interest and many applications since it was introduced by Aharonov, Albert and Vaidman in 1988 [1]; for some recent reviews see [2-7].

One important application of the concept is to more basic questions in quantum mechanics (QM); some of these aspects are reviewed, *e.g.,* in [2]. More recent examples are Vaidman's criterion for the past of a quantum particle [8] and Aharonov's and his collaborators' focus on aspects of QM correlations in composite systems [9,10]. Such uses of weak values have also met with some criticisms, *e.g.*, in [11-14]. In all these cases the issue is very much concentrated on what *property* of the system under investigation a weak value represents, what *meaning* one should assign to it, in other words how a weak value is to be *interpreted in physical terms.*

In this paper, I study one particular aspect of this interpretational issue. In particular, I investigate some conditions that have to be fulfilled in order for the weak value at all to represent the system under study. Indeed, it turns out that the weak measurement involved in extracting a weak value, despite it being a *weak* measurement, in some cases – to be specified below – influences the system to such an extent that the relevant system state after the measurement will be Hilbert space orthogonal to the undisturbed system state. Neither this after-measurement system state, nor the weak value constructed from it, can therefore represent the system *per se*.

The paper starts by giving the necessary general background on which my arguments are based. I then discuss the condition for the weak measurement to result in a weak value representing the system under investigation; it hinges on the fit to the no-measurement case of the limit when the weak measurement strength tends to zero. Special attention has to be paid to the case when the system evolves between the weak measurement and the postselection. A few examples will illustrate the main findings; in particular, I find that the use of the weak value by Vaidman in [8] and by Aharonov and collaborators in [9] violates this condition. The weak values they use can, therefore, not be considered as representing a property of the system they investigate.

## II. BASIC FORMALISM

In quantum mechanics, the *weak value* $\mathcal{S}_w$ of a an operator $\mathcal{S}$ for a quantum system $\mathfrak{S}$ has the formal definition (in the simplest cases; see eq (7) below for a more general expression for $\mathcal{S}_w$)

$$\mathcal{S}_w = <f|\mathcal{S}|in>/<f|in> \qquad (1)$$

where $|in>$ is the initial, "preselected", state and $|f>$ is the final, "postselected", state for the system, and where $<f|in>$ is assumed not to vanish..

As such, the weak value (1) is the quotient of two amplitudes [11]. It is therefore not immediately clear what this entity represents physically, *i.e.,* how one should *interpret* the expression (1) in more physical terms, in particular what a particular numerical value should mean. The basic rules of QM have nothing to say on this matter [12,13]. True, attempts have been made to relate $\mathcal{S}_w$ to a (complex) probability – see, *e.g.*, [15] and references therein – but this requires an extension of the usual concepts of a probability. Here, I shall not dwell on such approaches but stick to conventional QM.

Another way to approach the meaning of a weak value is via its operational definition, *i.e.*, to relate it to the way it is actually measured. In the usual treatment [1-7] one considers the interaction between the system $\mathfrak{S}$ under study and a measuring device, a meter $\mathfrak{M}$, assuming a von Neumann-type measurement interaction expressed by the unitary operator

$$U_M(S) = \exp(-i g\, S \otimes P_M).\tag{2}$$

Here, the meter is schematized as a quantum system described by the "pointer variable" $Q_M$ – the eigenvalues of which are denoted $q$ – and its conjugate momentum $P_M$. Furthermore, $g$ is a measure of the interaction strength and the symbol $\otimes$ stands for the direct product of the system and meter Hilbert spaces.

Working in the interaction picture (or the Schrödinger picture with all free Hamiltonians put to zero), the procedure to arrive at (the real part[1] of) a weak value consists of the following steps (see [2-7] for more detailed expositions with all the necessary details and caveats):

(a) Prepare the joint system-meter setup in the state $|in> \otimes |m>$, with $|m>$ the initial state for the meter (assumed, for simplicity, to have a Gaussian wave function centered at zero and with width $\Delta$ in its square).

(b) Let the system and the meter interact according to (2), meaning a transition to the state $U_M(S)\,(|in> \otimes |m>)$; I will more precisely refer to this transition as a *(pre-)measurement* and also say that a weak *meter probe is inserted* at this stage of the evolution.

(c) As an option, allow the system to undergo further, internal (unitary) evolutions, *i.e.*, evolutions not involving the meter.

(d) "Postselect" the state $|f>$ for the system, *i.e.*, make a projective measurement on the system of some observable $F$ and consider only those events that give a particular eigenvalue $f$ of $F$, with corresponding eigenstate $|f>$, assumed for simplicity to be nondegenerate.

(e) After this postselection, subject the meter to a projective measurement of its pointer variable $Q_M$, giving a result $q$.

(f) Repeat the procedure to get enough $q$-statistics.

(g) Obtain the mean value $_f<Q_M>$ of the $q$-distribution.

(h) Establish the weak value from the relation[2]

$$S_w = \lim_{g \to 0} [\,(1/g)\ _f<Q_M>\,].\tag{3}$$

---

[1] The imaginary part of the weak value may be obtained from a similar procedure measuring the momentum observable $P_M$ for the meter. For simplicity, I shall assume throughout this paper that the weak value is real. In discussing matters of principles as I do, this is no essential limitation.

[2] Aharonov, Vaidman and their collaborators prefer to use the limit $\Delta \to \infty$ as the weak measurement limit, see, *e.g.*, [2]. This is equivalent to the limit in (3) but has the drawback that it cannot easily be used to discuss the $g = 0$ case.

This scheme gives (3) as the operational meaning of the weak value $S_w$. Still, it does not immediately furnish an interpretation of $S_w$ [11-13]. In particular, it does not tell what a particular numerical value of $S_w$ might mean.

However, in some applications it is rather the vanishing or not of the weak value that is the interesting question. For example, Vaidman [8] has proposed a "weak trace" criterion for the presence of a particle. His purpose is to find a way of defining what is to be meant by "the past of a QM particle" *i.e.*, to be able to say what path one might assign to a particle in a given setup with several "channels" open for the particle (I shall give a more formal definition of my notion of a "channel" in connection with eq. (8) below but a typical example is a photon in a Mach-Zehnder interferometer (MZI) with the arms as the "channels"). Vaidman proposes to formulate such a criterion in terms of the weak value of the projection operator onto that part of the setup where one wants to establish the presence or not of the particle: a nonvanishing such weak value would then be a sign that the particle has been present at that location. This "weak trace" criterion has also been invoked, more or less explicitly, in several other applications of the weak measurement approach (see for example [9,10,16]).

Disregarding some objections (shortly to be specified) that could be raised to such notion of the "presence" of the system in a particular "channel", the Vaidman criterion is basically a sound one: a weak value of a projection operator can be non-zero only if there is a contribution to the total state of the system under study from the particular "channel" interrogated by the (pre-)measurement of the corresponding projection operator.

One objection has to do with the fact that a criterion formulated in terms of the weak value crucially depends also on the choice of the final, postselected state. And this crucial dependence on the postselected state could, in fact, distort the weak value and masks the exposition of the particular channel one wants to focus on. An (extreme?) example is when the denominator $<f \mid in>$ of $S_w$ in (1) happen to vanish; here, I shall assume this not to be the case.

Another objection, which will be my focus in the rest of this paper, is the following. The operational definition (3) of the weak value involves the limit $g \to 0$. As always when one works with limits, if one is to use this limit as representing the $g = 0$ case, one must convince oneself that the limit is continuous. In fact, this is *not* the case in some of the well-studied applications of the weak value approach. As I will show, the operational definition (3) of the weak value could namely be discontinuous in the sense that the limit $g \to 0$ does not represent the undisturbed $g = 0$ case. The reason is that the (pre-)measurement, despite it being weak, disturbs the system to such extent that the result of taking the limit $g \to 0$ cannot any longer be said to represents the undisturbed system. In these cases, the weak value therefore loses its role as a characteristic of the undisturbed system.

## III. CONDITIONS FOR CONTINUITY

### III.1 General considerations

To investigate whether the $g \to 0$ limit joins smoothly to the $g = 0$ case, consider the time evolution of the initial state due to the (pre-)measurement interaction of point (b) above[3]

$$| in > \otimes |m> \; \to \; U_M(\mathcal{S})\,(| in > \otimes |m>) = exp\,(-i\,g\,\mathcal{S} \otimes P_M)\,(| in > \otimes |m>) \quad (4)$$

To linear order in the interaction strength $g$, this reads

$$U_M(\mathcal{S})\,(| in > \otimes |m>) \approx | in > \otimes |m> - i\,g\,\mathcal{S}| in > \otimes P_M |m> \quad (5)$$

It is the factors $-i\,g\,\mathcal{S}| in >$ in the second term – or rather this factor divided by $g$ as implied by (3) – that is the source of the weak value.

To allow for point (c) in the weak measurement scheme outlined above, suppose that the system undergoes further unitary evolutions after the interaction with the meter but before postselection. Let me for simplicity assemble these evolutions in one unitary operator $U_{syst}$. Then, the postselection implies that the (non-normalized) state of the meter, still to linear order in the interaction strength $g$, becomes

$$<f|\,U_{syst}\,U_M(\mathcal{S})\,(| in > \otimes |m>) \approx <f|\,U_{syst}\,[| in > \otimes |m> - i\,g\,\mathcal{S}| in > \otimes P_M |m>] =$$

$$= <f|\,U_{syst}\,| in > \,[1 - i\,g\,\mathcal{S}_w\,P_M)]\,|m> \approx$$

$$\approx <f|\,U_{syst}\,| in > \,exp\,(-i\,g\,\mathcal{S}_w\,P_M)\,|m>, \quad (6)$$

where, compared to (1), the weak value now takes the slightly more general form

$$\mathcal{S}_w = <f|\,U_{syst}\,\mathcal{S}\,| in >/<f|\,U_{syst}\,| in >, \quad (7)$$

where as usual I assume that $<f|\,U_{syst}\,| in > \neq 0$.

The expression (6), for $U_{syst} = 1$, gives one way of understanding the result (3): the meter wave function is shifted by the amount $g\,\mathcal{S}_w$ with respect to the $g = 0$ case.

### III.2 No subsequent system evolution.

Firstly, let me consider the simpler case with $U_{syst} = 1$, *i.e.*, no further evolution of the system after the weak (pre-)measurement, so that the original expressions (1) and (3) apply.

---

[3] One should be aware of the fact that the state $|in>$ is the state of the system immediately before the system-meter (pre-)measurement interaction.

Then, one of two things may happen. The first is that $< in \mid \mathcal{S} \mid in > \neq 0$, so the state $\mathcal{S} \mid in >$ has a component along $\mid in >$. The term involving the factor $i\,g\,\mathcal{S} \mid in >$ in the expression (5) then just appears as a (small) term added to an already existing (large) term in the undisturbed system state $\mid in >$. The limit $g \rightarrow 0$ is then a smooth one and the weak value may be considered a true representative of the undisturbed system.

To illustrate, let $\mid in >$ have an expansion

$$\mid in > \; = \mid a > + \mid b > + \mid c > + \ldots \ldots \tag{8}$$

in terms of orthogonal (but not necessarily normalized) state vectors $\mid a >, \mid b >, \mid c >,$ *etc.*, each representing what I call a "channel" for the undisturbed system. The condition $< in \mid \mathcal{S} \mid in > \neq 0$ is obeyed, *e.g.*, for $\mathcal{S} = \Pi_a = \mid a >< a \mid / < a \mid a >$, the projection operator onto one "channel" $\mid a >$ of the system. The smoothness of the limit $g \rightarrow 0$ is obvious.

The second case is that $< in \mid \mathcal{S} \mid in > \; = \; 0$. Then there is a problem: the state $\mathcal{S} \mid in >$ is orthogonal to $\mid in >$ and has no component along $\mid in >$. In fact, the weak (pre-)measurement interaction disturbs the system to such extent that the state $g\,\mathcal{S} \mid in>$ appearing in the weak value source term in (5) and (6) is orthogonal to the state of the undisturbed system. The (pre-)measurement so to speak "derails" the system away from its undisturbed state into an orthogonal state. The fact that this "derailment" is "tiny" [8] is not relevant: the smallness is entirely due to the factor $g$, which is divided out in (3) to get the weak value[4]. And the fact that a state is in an orthogonal subspace is a discrete fact in the sense that it cannot be changed continuously: the state $g\,\mathcal{S} \mid in>$ being in an orthogonal subspace – *i.e.* orthogonal to the undisturbed system state – for any non-vanishing $g$ means that this situation prevails in the limit $g \rightarrow 0$.

The conclusion in this case is that the limit in (3) involves a state, *viz.*, $\mathcal{S} \mid in >$, that does not describe the (undisturbed) system: when $< in \mid \mathcal{S} \mid in > \; = \; 0$, the weak value does not represent the undisturbed system. (The case $< in \mid \mathcal{S} \mid in > \; = 0$ due to $\mathcal{S} \mid in > \; = 0$ requires special attention and is commented on in Appendix A.) It follows that the weak value in this case is constructed from a state that does not describe the undisturbed system and therefore does not describe any (physical) property of the undisturbed system either. It is rather an artefact of the system-meter (pre-)measurement interaction. It cannot be used to physically characterize the system.

---

[4] An analogy to this situation, although not a perfect one, is an ordinary function *f(x)* which for small *x > 0* has an expansion *f(x) = a x + O(x²)* but which for *x > 0* has a non-vanishing value. Then one has *lim$_{x \rightarrow 0}$* [(1/x) *f(x)*] = *a,* while for *x = 0* the expression $\frac{1}{x}$ *f(x)* is undefined (or, if you prefer, infinite). – See also the summary and conclusion section below.

In sum, as long as (i) there is no further (internal) evolution of the system between the weak (pre-)measurement) and the postselection, *i.e.*, as long as $U_{syst} = 1$, and (ii) the observable $\mathcal{S}$ under study obeys the condition $< in \,|\, \mathcal{S} \,|\, in > \neq 0$, then the weak value (2) does characterize the system. In the sequel, I shall call such weak value "well-behaved". What kind of physical property this is in general is, however, not established by my analysis. In the special case of $\mathcal{S}$ being a projection operator, the vanishing or not of its weak value may, though, be linked – following Vaidman [8] as outlined above – to the presence or not of a particular "channel" for the system in the preselected state $|\, in >$.

If, on the contrary, $< in \,|\, \mathcal{S} \,|\, in > = 0$, then the weak value cannot be used at all to characterize the system.

## III.3 Examples with no subsequent system evolution

Let us see how these ideas work in concrete cases.

*Example 1 : A simple Mach-Zehnder device*

Let in this case the system be a photon passing through a simple, well-balanced Mach-Zehnder interferometer (MZI) as illustrated in Fig. 1. This figure also gives my notation. In particular, I denote the state of the photon in the left, *L*-arm by $|\, L >$, *etc*; these states in the different arms of the MZI are examples of what I call "channels". One wants to establish whether the photon could have passed the *L*-arm, say. With Vaidman's "weak trace" criterion [8], one should consider a weak value of the projection operator $\Pi_L = |\, L > < L \,|$ onto the left arm of the MZI. With the preselected state (the state of the system immediately before the weak measurement interaction) given by

$$|\, in > = (\,|\, L > + \, i \,|\, R >\,) / \sqrt{2} \,, \tag{9}$$

it trivially follows that

$$< in \,|\, \Pi_L \,|\, in > \neq 0. \tag{10}$$

Furthermore, let $|\, f >$ be any of the states $|\, L´ >$ or $|\, R´ >$ for the photon after it has passed the second beamsplitter. True, this involves what I have called a further system evolution, but here this evolution is rather harmless, so it may be disregarded. The conclusion is then that the weak value of $\Pi_L$ is representative of the non-disturbed system and since it is non-vanishing, it establishes, to nobody's surprise, that photon could have passed the *L*-arm of the MZI before it gets postselected.

*Example 2: The Quantum Cheshire cat.*

In brief outline, the authors of [9] consider an MZI with polarized photons (but polarization-insensitive beamsplitters) and a preselected state (the state of the system just before the system-meter (pre-)measurement interaction) essentially given by

$$|in> = (|R>|V> + |L>|H>)/\sqrt{2}. \qquad (11)$$

Here, $|V>$ ($|H>$) is the vertical (respectively the horizontal) polarization state for the photon.

As the postselected state, the authors chose

$$|f> = |L'>|H>. \qquad (12)$$

If one is only interested in which path the photon could have taken, not bothering about its polarization, analogous arguments to those for the simple MZI in example 1 above, result in the conclusion that the projectors $\Pi_L$ and $\Pi_R$ are both "well-behaved". With the preselected state given by (11) and the postselected by (12), it follows that the weak value of $\Pi_L$ is non-vanishing, while that for $\Pi_R$ is zero. As the authors phrase it: The cat (*alias* the photon) is in the left arm, not in the right one.

With the same pre- and postselected states as above – (11) and (12) respectively – the authors then look at the weak value of

$$\sigma_z^{(R)} = \Pi_R \otimes \sigma_z \quad , \qquad (13)$$

and of an analogous expression for $\sigma_z^{(L)}$. Here, in terms of the circular-polarization eigenstates $|\pm> = (|V> \pm i|H>)/\sqrt{2}$, the photon circular polarization (or "spin") is

$$\sigma_z = |+><+| - |-><-| . \qquad (14)$$

In [9], this observable is identified with the "grin" of the quantum Cheshire cat.

Now, (except for an overall factor of *i* and a possible overall sign)

$$\sigma_z \sim |V><H| - |H><V| . \qquad (15)$$

Then, one straight-forwardly deduce that the weak value of $\sigma_z^{(R)}$ is non-zero while that of $\sigma_z^{(L)}$ vanishes, *i.e.,* with the presence in the *L* and *R* arms interchanged compared to the no-spin case above. The authors conclude that the grin is where the cat is not, a true Cheshire cat behavior.

There is, however, one objection. It is easy to see that the operators under study, the essential factor of which is the expression (15), are not "well-behaved" in the terminology introduces above: the matrix elements $<in|\sigma_z^{(R)}|in>$ and $<in|\sigma_z^{(L)}|in>$ vanish, meaning that the operator $\sigma_z^{(R)}$ (respectively $\sigma_z^{(L)}$) "derails" the system away from its undisturbed state. Consequently, any conclusion drawn in [9] regarding the possible location of the "grin" cannot be upheld: the weak value of (13) cannot be used to physically characterized the undisturbed system.

*III.4 Allowing for subsequent system evolution.*

Let me next consider the case when there is a unitary system evolution in between the system-meter (pre-)measurement interaction and the postselection, *i.e.*, when $U_{syst}$ in (6) is not the unit operator.

In this case, and contrary to what was possible for the $U_{syst} = 1$ case treated previously, it is difficult to find a concisely formulated general condition for the weak value really to represent the system. However, a basic observation is that, in order for a weak value (7) to be "well-behaved", *i.e.*, for it really to characterize the undisturbed system, a necessary condition is the following: at no moment in its subsequent evolution shall the state evolving from $\mathcal{S}|in>$ by $U_{syst}$ permanently "derail", *i.e.*, shall not permanently evolve into a state orthogonal to the state of the undisturbed system.

*III.5 Example with subsequent system evolution.*

Instead of trying to formulate a general criterion, let me give one example of what may occur.

*Example 3. Two nested MZIs.*

This example closely follows the presentation in Vaidman [8]. In particular, and referring to my Fig. 2, the photon state in arm *A* is labelled $|A>$, *etc*. Among other things, Vaidman is interested in establishing the presence or not in the *B*-arm of those photons that end up in the detector $D_2$. It is also stipulated that the inner MZI, the one with arms *B* and *C*, is so tuned that, for the undisturbed system, no photons from this MZI enter into arm *E*. This is accomplished by having suitable interference between the *B*- and *C*-arm photonic states in the beamsplitter *BS3* so that, for the undisturbed system, the photons in the innermost MZI all end up in the detector $D_3$.

The state to be chosen as the preselected state is the one just before the photon reaches the meter probe inserted in the *B*-arm. With my conventions for transitions through a well-balanced beamsplitter (see Appendix B), it reads

$$|in> = (i\sqrt{2}|A> + i|B> + |C>)/2 . \qquad (16)$$

With the system-meter interaction, to linear order in *g*, given by

$$U_M(\Pi_B) = exp(-ig\,\Pi_B \otimes P_M) \approx 1 - ig\,\Pi_B \otimes P_M , \qquad (17)$$

the joint system-meter state right after this interaction is given by

$$U_M(\Pi_B)(|in> \otimes |m>) \approx |in> \otimes |m> - ig\,|B> \otimes P_M\,|m>. \qquad (18)$$

Let us now see what happens when the system evolves further through the beamsplitters *BS3* and *BS4*; the meter state is not affected, so I can leave that aside for the moment. Since the MZI setup is so arranged that there is complete interference in *BS3* beamsplitter in favor of the *D₃* detector arm, the undisturbed system, represented by the first term on the right hand side of (18), evolves into

$$U_{syst}(BS3)\,|\,in> \;=\; i\;(|\,A> + |\,D_3>)/\sqrt{2}. \tag{19}$$

Note, in particular, that there is no *E*-arm state here.

The situation is different for the second term in (18), the one induced by the (pre-)measurement and involving the state $g\,|\,B>$. Since here the state $|\,B>$ occurs alone, without any compensating *C*-arm state to interfere with, it does evolve into a state with a component into the *E*-arm:

$$U_{syst}(BS3)\,|\,B> \;=\; (|\,D_3> + i\,|\,E>)/\sqrt{2}\,. \tag{20}$$

In the terminology introduced above, and disregarding the term $|\,D_3>$ – it is of no importance for a weak value involving the postselected state $|\,D_2>$ – one sees that the weak system-meter interaction $U_M(\Pi_B)$ "derails" the system from the state it occupies when undisturbed and into an orthogonal state. It is this "derailed" term, more precisely the state $|\,E>$ in (20), that via the beamsplitter *BS4* can allow the photon to reach the postselection of $|\,f> = |\,D_2>$ and there be the source of the (non-vanishing) weak value of the projector $\Pi_B$ onto the *B*-arm. But a photon undisturbed by the (pre-)measurement evolves according to (19). The weak value (7) is in this case just a result of the system-meter interaction and does not represent the undisturbed system.

Vaidman's conclusion from the non-vanishing of this weak value is that the photons in the undisturbed state may be present in the *B*-arm and anyhow reach the detector *D₂* without passing the *E*-arm. In my opinion, this conclusion cannot be upheld: even if ever so weak, the system-meter interaction used to establish this weak value disturbs the system to such an extent that it "derails" the system from its undisturbed state. Expressed in other words: As long as the (pre-)measurement strength *g* is non-zero, the answer to the question "Could there be photons in the arm *E*?" is "yes". But for *g* = 0, the answer to the same question is "no". The weak value under consideration does not characterize the undisturbed system.

## V. SUMMARY AND CONCLUSION

In this paper, I have investigated some questions related to the interpretation of a weak value for a quantum mechanical system $\mathbb{S}$. In particular, I have shown that there are certain conditions to be obeyed in order for the weak value at all to represent the system that it is supposed to characterize. These conditions depend on the way the weak (pre-)measurement system-meter interaction influences the system. More precisely, if this interaction leaves the system state with a component along the undisturbed system state, then the weak value does characterize the system *per se.* If, on the other hand, the (pre-)measurement interaction takes the state of system $\mathbb{S}$ entirely into a state that is orthogonal to the state that $\mathbb{S}$ occupies when undisturbed, then the corresponding weak value does not represent the undisturbed system $\mathbb{S}$ but is just an artifact of the system-meter (pre-)measurement interaction. In a shorthand wording, I have described this latter situation by saying that the (pre-)measurement "derails" the system into a state orthogonal to the undisturbed system state.

It is of no importance for this argument that the (pre-)measurement is weak as long as there is a (pre-)measurement interaction at all, *i.e.*, as long as the parameter *g* of eq. (2) does not vanish. This is so because the strength *g* of the interaction is divided out when one deduces the weak value from the relation (3). In other words: the question whether the weak value generating state $g \, \mathcal{S} \, | \, in >$ lies entirely in an orthogonal subspace or not can be given a yes or no answer for all nonvanishing values of *g*. And if the answer is yes, then this state does not represent $\mathcal{S}$, implying that neither does the weak value: it is constructed from a state that is orthogonal to the state of the undisturbed system and does not describe $\mathcal{S}$.

Of course, it is no question of a discontinuity of the type that occurs in the "collapse" of the state as it occurs in the usual, orthodox QM description of a projective measurement. It is instead a question of the weak (pre-)measurement procedure to (weakly) influence the system so that its state is orthogonal to the undisturbed system state. But it is this "derailed" state – after division by *g* (an action which so to speak gets rid of the weakness; compare footnote [4]) above) – that gives the weak value, which is thus constructed from a state not describing the undisturbed system. So the situation is maybe better characterized as an example of a topological discontinuity, somewhat like a bifurcation as described in so called catastrophe theory.

On these arguments, conclusions reached in some applications which have been made of weak values cannot be upheld. I have given two examples: the so called quantum Cheshire cat of [9] and the nested Mach-Zehnder interferometers of [8].

**ACKNOWLEDGEMENT**

I am grateful to Ruth E Kastner for constructive comments to a preliminary version of this paper and to Bo Söderberg for several clarifying discussions.

**APPENDIX A: COMMENTS ON THE CASE $\mathcal{S}| \, in > \, = 0.$**[5]

The criterion $< in \, | \, \mathcal{S} \, | \, in > \, = \, 0$ is obeyed if $\mathcal{S} \, | \, in > \, = 0$, *i.e.* if the operator $\mathcal{S}$, whose weak value is under investigation, projects the preselected state $| \, in >$ onto the null Hilbert space. (Trivially, in this case the source term in (5) is zero, and so is therefore the weak value.) Whether or not, in the terminology introduced above, this should be characterized as a "derailment" – a projection away from the Hilbert space ray $|in>$ – cannot , I think, be decided on rational grounds. This is so because the null state is contained in any proper Hilbert subspace but is also orthogonal to all states. So whether a projection onto the null state should be said to project the state $|in>$ out of its original ray or not is more a matter of judgment. This situation is related to the maybe futile distinction between having a vanishing weak value representing the undisturbed system or not having a "well-behaved" weak value at all.

---
[5] I am grateful to E. Cohen for making me aware of the need for a special treatment of this case.

More to elucidate the question than to try to resolve it, let me here only consider one special case which has some application, *e.g.*, to the example treated in [16].

Consider a Hilbert space of at least three dimensions and suppose that the state $|in>$ has the expansion

$$|in> \ = |a> \ + |b> \qquad (A1)$$

in terms of orthogonal (but not necessarily normalized) state vectors $|a>$ and $|b>$. Further, let $\mathcal{S}$ be the projector

$$\mathcal{S} = \Pi_d \ = \ |d><d|/<d|d> \qquad A2)$$

onto a state $|d>$ orthogonal to both $|a>$ and $|b>$. Then, trivially,

$$\mathcal{S}|in> \ = 0 \ . \qquad (A3)$$

Although, as I said, the matter is up for discussion, I judge it appropriate, from the way this example is constructed – that the state $|d>$ is orthogonal to $|in>$ of (A1) – to say that it is here a question of "derailment" out of the ray (A1) of the undisturbed system.

**APPENDIX B : CONVENTION FOR TRANSITION IN A BEAMSPLITTER.**

In the notation of Fig.1, the transition in the first (well-balanced, 50-50) beamsplitter is in my convention described by the unitary transition

$$|N> \ \rightarrow \ U(BS1) \ |N> \ = (|L> \ + \ i|R>/\sqrt{2} \qquad (B1)$$

with analogous expressions for any other similar beamsplitting process considered in this paper.

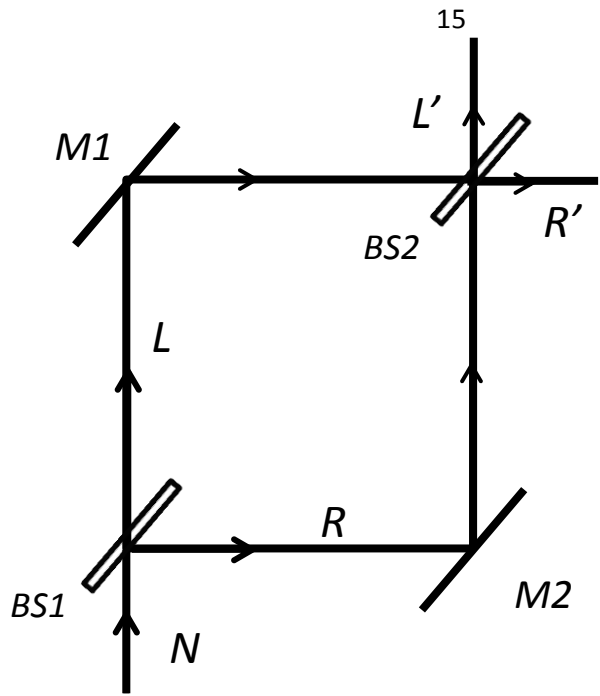

Figure 1. A simple Mach-Zehnder interferometer (MZI) setup, also exhibiting the notation used in the text. The beamsplitters are denoted *BS1, etc.*, the mirrors *M1, etc.*, and the arms *L, R , etc.*

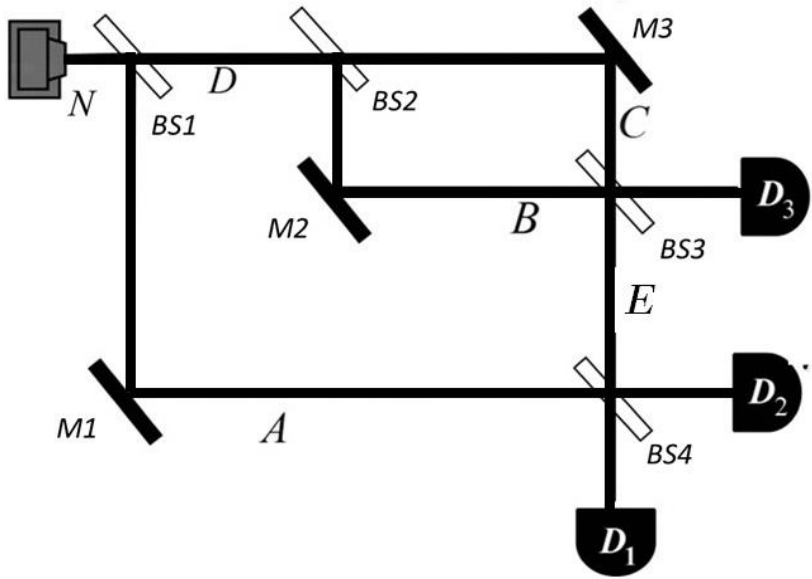

Figure 2. The nested MZI setup used in example 3, exhibiting the notation for the arms (*N, D, etc.*), mirrors (*M1, etc.*) and beamsplitters (*BS1, etc.*) used in the text (adopted from Vaidman [8] ),